\documentclass[twocolumn,showpacs,floatfix,superscriptaddress,amsmath,amssymb,prl]{revtex4}
\usepackage{mathrsfs}
\usepackage{amssymb}
\usepackage{graphicx}
\usepackage{hyperref}
\usepackage{ulem}
\usepackage{tabularx}
\usepackage{array}
\usepackage{color}
\begin{document}

\title{Quantum Probability assignment limited by relativistic causality}

\author{Yeong Deok Han}
\affiliation{Department of Information Security, Woosuk University, Wanju, Cheonbuk, 565-701, Korea}

\author{Taeseung Choi}
 \email{tschoi@swu.ac.kr}
\affiliation{Division of General Education, Seoul Women's University, Seoul 139-774, Korea}
\affiliation{School of Computational Sciences, Korea Institute for Advanced Study, Seoul 130-012, Korea}



\begin{abstract}
The quantum nonlocality is limited by relativistic causality, however, the reason is not 
fully understood yet.  
The relativistic causality condition on nonlocal correlations has been usually accepted as a prohibition of faster-than-light signaling, called no-signaling condition.  
We propose another causality condition from the observation that space-like separate events should have no causal relationship. 
It is proved that the new condition is stronger than no-signaling condition for a pair of binary devices. 
We derive the standard probability assignment rule, so-called Born rule, on quantum measurement, which determines the degree of quantum nonlocality, by using relativistic causality constraint.
This shows how the causality limits the upper bound of quantum nonlocality through quantum probability assignment.

\end{abstract}

\flushbottom
\maketitle
%
%
\thispagestyle{empty}


\section{Introduction}

The compatibility between nonlocality in quantum mechanics and causality in special relativity is a 
fundamental question but not fully understood yet. 
Quantum nonlocality manifests itself in the correlations of local measurement outcomes  
on an entangled state shared between two space-like separate parties, Alice and Bob \cite{Bell}. 
A local measurement on one state of entangled pair by Alice (Bob) influences the other state of Bob (Alice) instantaneously 
in the standard quantum physics because no physical processes are suggested in quantum measurement postulates. 
An important question is whether the hypothetical influence between the space-like separate pair 
can violate (relativistic) causality. 
The causality has been considered as a prohibition of faster-than-light signaling, which is 
called 'no-signaling' condition. 
Bancal et al. have shown that for any finite speed hypothetical influences, 
faster-than-light communication can be built \cite{Bancal}. 
According to their results, only when the speed of the hypothetical influence is infinite, 
the quantum nonlocality can be reconciled with no-signaling condition.  
Recent experiments determined that the lower bound of the speed of the hypothetical influence has to exceed 
the speed of light by at least four orders of magnitude \cite{Salart, Yin}. 
Those results suggest that the speed of the hypothetical influence would be infinite 
so that quantum nonlocality satisfies no-signaling condition.  

Another important question is whether quantum correlation gives the maximum nonlocality among the nonlocal 
correlations that satisfy the causality.  
The amount of nonlocality can be demonstrated by a violation of the Clauser-Horne-Shimony-Holt (CHSH) inequality, 
bounded by 2 in any local classical theory \cite{Clauser}.
The upper bound of quantum correlations, which is known as Tsirelson's bound, is 
$\mathcal{B}_Q=2\sqrt{2}$ \cite{Tsirelson}. 
Popescu and Rohrlich found the surprising result that the nonlocality of quantum correlation is not maximum 
and the maximum upper bound, which satisfies no-signaling condition, is 4 \cite{Popescue1}. 
As a result, they have shown the existence of 'superquantum' correlations that are more nonlocal than 
quantum correlations. 
Several attempts to explain the reason why post-quantum theory, which has superquantum correlations, 
was not found in nature have been proposed \cite{Brassard, Linden, Pawlowski, Navascues, Fritz, Navascues1}.
However, this problem is still an open question. 

Bell theorem \cite{Bell} implies that any nonlocal theory, which satisfies no-signaling condition, 
should not be deterministic. The indeterminacy of quantum theory is implemented by the probabilistic outcomes 
in the measurement through the collapse of a quantum state into an eigenstate of an observable. 
The measurement postulate in the standard quantum mechanics states that the probabilistic assignment to 
measurement outcomes is governed by Born rule \cite{Wilde}. 
At this stage we naturally ask a question: Can any nonlocal theory, which has superquantum correlations 
and preserves causality, be constructed just by imposing other probability assignment on quantum measurement 
than Born rule? 
As Pospescue and Rorhlich have shown, the change of probability assignments to correlations 
could provide more nonlocality than quantum correlations. Hence other probability assignment 
on quantum measurement than Born rule will provide new nonlocal correlations.
The issues are whether the exact form of quantum probability assignment can be found 
by imposing causality condition and it is Born rule. 
The no-signaling condition will be shown not tight enough to answer this question.  
Hence, the stronger condition as a causality constraint than no signaling condition is necessary 
to determine the explicit form of quantum probability assignment. 
We will propose new causality condition by the observation that the causality cannot be imposed between 
space-like separate events in special relativity. 
The new causality condition will be proved to be stronger than no-signaling condition between a pair of nonlocal 
binary devices. That is, the new causality condition is a sufficient but not a necessary condition 
for no-signaling condition between a pair of nonlocal binary devices. 
We will show that Born rule is the unique probability assignment rule on quantum measurement 
by using the new causality condition. 
This fact implies that a nonlocal theory with superquantum correlations cannot be obtained 
by the modification of quantum probability assignment from the standard quantum theory. 
  
\section{Results}

\subsection{Relation between two causality conditions}

First, we will establish the relationship between no-signaling condition and the new causality condition. 
The no-signaling condition, which requires that no signal can be transmitted faster than light, 
is usually accepted as the requirement of the causality 
because a faster-than-light signal could be used to send an information 
back to past in special relativity, that definitely violates causality.  
We propose another condition, we call 'no-causal-order' condition, as the requirement of causality 
from the observation that a causal order is based on temporal order, which requires that 
'cause' has to precede 'effect' for every observer. 
In special relativity, 
a causal relation of 'cause' and 'effect' could be established between two events, 
only when the two events are either light-like or time-like separated such that the time sequence of 
'cause' and 'effect' is absolute  because any Lorentz transformation cannot 
invert their time order.    
That is, the event 'effect' has to be either on the future 
light cone or inside the future light cone of the event 'cause'. And conversely, the event 
'cause' should be either on the past light cone or inside the past light cone of event 'effect'. 
However, the time sequence of any two space-like separated events could be changed according to 
the motion of an observer. This is because there is no absolute global time in special relativity, 
on which every observer agrees. 
Hence relativistic causality requires no causal relationship between a pair of space-like separated events. 
To our best knowledge, 
the constraint on the correlations of a pair of space-like separated events imposed by the requirement of 
no causal relationship did not formulated yet. 
The no-causal-order condition is motivated by this observation. 

We will define no-causal-order condition specifically for a pair of nonlocal devices with binary inputs and outputs 
shared by Alice and Bob. 
Here we assume that Alice and Bob are at rest with respect to each other for simplicity without loss of generality. 
An input and an output of Alice's device are $x$ and $a$, respectively, 
and $y$ and $b$ for Bob's device as in Fig. \ref{fig:NLDevices} in Methods. 
That is, the measurement outcome of Alice's (Bob's) device with input x (y) is a (b). 
The correlations between two devices are described by the joint probability distributions 
of outcomes $a$ and $b$ for inputs $x$ and $y$. 
We also assume that the measurement events of Alice and Bob are space-like separated. 
Let us consider an observer O who observes the operations (measurements) of Alice and Bob. In special relativity, 
even though there is no absolute global time, observer-dependent global time can be well defined. 
This means that for an observer O the temporal order of measurement events of Alice and Bob is clearly determined 
in O's own reference frame even though the two events are space-like separated. 
The following two possible temporal orders of the measurements of Alice and Bob for the observer O will be considered. 
First one is that Alice's measurement precedes Bob's measurement for the observer O, which we call Alice-first measurement. 
Second one is that the temporal order of Alice's and Bob's measurements is reverse for the observer O, 
which is called Bob-first measurement. 
A simultaneous measurements of Alice and Bob is not considered here because it is irrelevant to investigate the causality.  
Let the joint probabilities of outcomes $(a,b)$ for inputs $(x,y)$ 
in Alice-first (Bob-first) measurement for the observer O be 
$P_A(ab|xy)$ ($P_B(ab|xy)$) with the subscript A (B). 
To show different joint probabilities $P_A(ab|xy) \neq P_B(ab|xy)$, the two nonlocal devices operate by 
two different physical processes depending on the temporal order of two input operations. 
That is, the physical process $A$ operates in Alice-first measurement but the different physical 
process $B$ operates in Bob-first measurement. 
Let us consider another observer O$^\prime$ in another reference frame, who sees the reversed temporal sequence 
between Alice's and Bob's operation to the temporal order observed by O. 
When the observer O observes Alice-first measurement and the process $A$,  
the observer O$^\prime$ observes Bob-first measurement and the process $B$. 
This requires the equality of joint probabilities $P_A(ab|xy)$ and $P_B(ab|xy)$ 
because the physical process is independent on the reference frame in special relativity. 
This equality implies that Alice-first and Bob-first measurements could not be distinguished so that no causal relation 
can be established between two space-like separated events of Alice and Bob.
Therefore, we define the so-called 'no-causal-order condition' between the joint probability distributions of 
two space-like separated operations of nonlocal devices as 
\begin{eqnarray}
\label{eq:NCO}
P_A(ab|xy)=P_B(ab|xy)=P(ab|xy).
\end{eqnarray} 
Notice that the usage of different subscripts 'A' and 'B' in joint probabilities 
$P_A(ab|xy)$ and $P_B(ab|xy)$ according to the temporal 
order of measurements are necessary even though no-causal-order condition requires that they are equal, 
because it is natural for nonlocal devices to show different input-output correlations depending 
on the temporal order of operations and the equality of two joint probabilities $P_A(ab|xy)$ and $P_B(ab|xy)$ 
are required by another physical context, relativistic causality. 
Notice also that this no-causal relationship includes the local measurement condition that the result of 
a local measurement at one position must solely determined by local information at that position. 
The local measurement condition is written as $P_A(a|x)=P_B(a|x)$. 
This condition can be rewritten as $P_A(a|x)=\sum_{b,y} P_A(ab|xy)/2=P_B(a|x)=\sum_{b,y}P_B(ab|xy)/2$ 
by using our joint probability notation. Then one can easily checked that no-causal-order condition leads the 
locality condition. 


Now we will investigate the relations between no-signaling and no-causal-order conditions. 
To study the relations, no-signaling condition has to be rewritten with the subscripts 'A' and 'B' to 
denote whose measurement is the first one, because in our treatment the joint probabilities for a pair of events 
would be different according to the temporal order of the events until no-causal-order condition is invoked. 
No-signaling condition requires that the probabilities of outcomes of a receiver must be the same 
independent on the choice of inputs of a sender. 
Let us suppose that Alice (sender) tries to send a signal to Bob (receiver), then no-signaling condition 
in our denotation becomes 
\begin{eqnarray}
\label{eq:NS}
&&P_A(b|0y)=\sum_{a\in \{0,1\}} P_A(ab|0y) \\ \nonumber
&=& \sum_{a\in \{0,1\}} P_A(ab|1y)=P_A(b|1y),
\end{eqnarray} 
for $b$, $y \in \{0,1\}$. 
$P_A(b|xy)$ is the marginal probability of Bob's input $y$ and outcome $b$ for Alice's input $x$ and 
the subscript $A$ denotes Alice-first measurement. 
The above formula implies that the probability of measurement outcome $b$ of Bob with his input $y$ is 
not dependent on whether Alice's input is either $0$ or $1$. 
Hence if Bob is a sender, the no-signaling condition becomes 
$P_B(a|x0)=P_B(a|x1)$, for $a$, $x \in \{0,1\}$. 
We will show that no-causal-order condition is stronger than no-signaling condition between a pair of 
binary nonlocal devices by proving the following proposition.

{\it Proposition}: Between a pair of nonlocal binary devices, no-causal-order condition $P_A(ab|xy)=P_B(ab|xy)$ is 
a sufficient, but not a necessary, condition for no-signaling condition $P_A(b|0y)=P_A(b|1y) $ 
and $P_B(a|x0)=P_B(a|x1)$.

{\it Proof.}- The proposition implies whenever no-causal-order condition is satisfied, no-signaling condition 
will also be satisfied (sufficient), but even if no-signaling condition is satisfied, no-causal-order condition will 
not always be satisfied (not necessary). We will first prove the sufficient condition, which can be 
rewritten with the following logically equivalent proposition (contrapositive): If no-signaling condition is not satisfied, 
then no-causal-order condition cannot be satisfied. 
Since it is assumed that no-signaling condition is not satisfied, 
a superluminal signal can be sent from a sender to a receiver. 
Let us suppose that Alice and Bob are a sender and a receiver, respectively, for clarity. 
Alice is able to send a superluminal signal to Bob by using her choice of inputs of her measurement in 
Alice-first measurement. 
After Alice's measurement, Bob can receive a signal, which is encoded in Alice's input, by measuring 
the different probabilities for his input $y$ and outcome $b$ depending on Alice's inputs, i.e., 
by using $P_A(b|0y) \neq P_A(b|1y)$. 
Now we are supposed to prove that this breakdown of no-signaling condition will lead to the breakdown 
of the no-causal-order condition. 
We will prove it by contradiction: First, we assume that no-causal-order condition is satisfied, and then show 
that the consequences of this assumption will give a contradiction.  
The no-causal-order condition requires that the probability distributions of Alice-first and Bob-first measurements 
are the same, that is, 
$P_A(b|0y)$$=$$ P_B(b|0y)$ and $P_A(b|1y)$$=$$P_B(b|1y)$. 
This condition derives the inequlity, $P_B(b|0y) \neq P_B(b|1y)$ in Bob-first measurement, which implies that 
the probabilities of Bob's measurement with the input and output pair $(y,b)$ depend on 
whether Alice's input is $0$ or $1$. In Bob-first measurement, Alice's inputs are not determined yet, so 
$P_B(b|0y) \neq P_B(b|1y)$ implies that Alice can send a superluminal signal from future to past. 
The possibility of a bi-directional signaling in time gives a contradiction (Appendix). 
Hence no-causal-order condition should not be satisfied when
no-signaling condition is not satisfied. 
We have proved that no-causal-order condition is a sufficient condition for no-signaling condition. 

Next we will prove that no-causal-order condition is not a necessary condition for no-signaling condition. 
We will prove this statement by providing an example, in which the joint probability distributions 
satisfy no-signaling condition but not no-causal-order condition. 
Let us consider an example of the nonlocal binary devices, which have the joint probabilities for input and output 
pairs of Alice and Bob as in Table \ref{tab:PRAntiPR}: 
The devices operate as the PR box in Alice-first measurement such that the joint probabilities are
$P_A(ab|xy)=1/2$ if $a\oplus b=xy$, and $0$ otherwise. Here $\oplus$ denotes an addition modulo $2$. 
The same devices show anti-PR box correlations in Bob-first measurement, defined by 
$P_A(ab|xy)=0$ if $a\oplus b=xy$, and $1/2$ otherwise. These devices definitely violate no-causal-order 
condition because $P_A(ab|xy)\neq P_B(ab|xy) $. 
It is easy to check that the joint probability distributions of PR box and anti-PR box in Table \ref{tab:PRAntiPR} 
both satisfy no-signaling conditions that are $P_A(b|xy) =P_A(b|\tilde{x} y)$ and 
$P_B(a|xy) =P_B(a|x \tilde{y})$. 
The tilde operation is defined as $\tilde{x}=x+1$ and $\tilde{y}=y+1$ modulo $2$. 
The above arguments prove that no-causal-order condition is not a necessary condition for no-signaling condition. 
Q.E.D.\\ 
According to the {\it Proposition}, no-causal-order condition is stronger than no-signaling condition 
between a pair of nonlocal binary devices. Hence we will 
impose no-causal-order condition on probability assignments to quantum measurement of a pair of qubits 
as the requirement of relativistic causality.

\begin{table}[ht]
\centering
\begin{tabular}{|l|l|l|l|c|c|}
\hline
a & b & x & y & $P_A(ab|xy)$ & $P_B(ab|xy)$\\
\hline
0 & 0 & 0 & 0 & $\frac{1}{2}$ & 0 \\
0 & 1 & 0 & 0 & 0 & $\frac{1}{2}$\\
1 & 0 & 0 & 0 & 0 & $\frac{1}{2}$\\
1 & 1 & 0 & 0 & $\frac{1}{2}$ & 0\\
\hline
0 & 0 & 0 &  1 & $\frac{1}{2}$ & 0\\
0 & 1 & 0 & 1 & 0 & $\frac{1}{2}$\\
1 & 0 & 0 &1 & 0 & $\frac{1}{2}$\\
1 & 1 & 0 & 1 & $\frac{1}{2}$ & 0\\
\hline
0&0& 1& 0 & $\frac{1}{2}$ & 0\\
0&1&1&0&0 & $\frac{1}{2}$\\
1&0&1&0&0 & $\frac{1}{2}$\\
1&1&1&0& $\frac{1}{2}$ & 0\\
\hline
0&0&1&1&0 & $\frac{1}{2}$\\
0&1&1&1& $\frac{1}{2}$ & 0\\
1&0&1&1& $\frac{1}{2}$ & 0\\
1&1&1&1&0 & $\frac{1}{2}$\\
\hline
\end{tabular}
\caption{Probability distributions for Alice-first and Bob-first measurement 
is determined by those of PR box and Anti-PR box, respectively.}
\label{tab:PRAntiPR}
\end{table}

\subsection{Derivation of Born rule}
 
The nonlocality of standard quantum physics is bounded by $2\sqrt{2}$ in CHSH correlation. 
Let us investigate whether new nonlocal theories with superquantum correlations could be obtained by 
generalizing the quantum probability assignment, given by the Born rule in standard quantum physics, 
under the requirement of relativistic causality, i.e., no-causal-order condition.  
In the standard quantum framework, a physical observable $\mathcal{A}$ is a linear Hermitian operator with 
real eigenvalues $\{ a_1, \cdots, a_d\}$ 
and mutually orthonormal eigenvectors $\{ |\phi_1\rangle, \cdots, |\phi_d\rangle\}$, 
where $d$ is the dimension of a Hilbert space. 
Then a general quantum state $|\psi\rangle$ is represented as a linear superposition of eigenstates. 
Physical observables satisfy the following measurement postulates: i) an outcome of a measurement is always an eigenvalue of 
$\mathcal{A}$. ii) The probability of an outcome $a_k$ for the initial state $|\psi\rangle$ is obtained 
with $f(a_k)=|\langle \phi_k|\psi\rangle|^2$.
iii) The quantum state after the measurement that gives the outcome $a_k$ 
reduces to the corresponding eigenstate $|\phi_k\rangle$.
The modification of i) has nothing to do with nonlocal correlations because correlations are implemented 
by an outcome probability not by the value of an outcome itself.  
The modification of iii) is not desirable because it is natural for physical systems that 
sequential measurements without any perturbation would give 
the same measurement results for the same observable $\mathcal{A}$.
Hence we will focus on a possible generalization of the quantum probability assignment of
postulate ii), which is known as Born rule, under the constraint of no-causal-order condition.

Let us define the generalized probability assignment rule on a state in a two-dimensional Hilbert space. 
We consider a qubit, which is given by 
the state $|\phi\rangle = c|0\rangle_z +d |1\rangle_z$, where 
$|c|^2+|d|^2=1$ and $|0\rangle_z$ and $|1\rangle_z$ denote eigenvectors of an input (observable) 
$z$ which has eigenvalues $0$ and $1$, respectively. 
For our purpose, it is enough to consider a pure state, 
because the mixed state is just a statistical mixture of pure states.  
The probability assignment for quantum measurement on the state $|\phi\rangle$ can be generalized from Born rule 
by applying an arbitrary non-negative real function $H(c)$ of complex number $c$ 
to the measurement probability $P(0|z)$ of the outcome $0$ for the input $z$, i.e., $H(c)=P(0|z)$. 
The other measurement probability $P(1|z)$ for the same input $z$ and the output $1$ 
can be determined by the normalization of probability as $P(1|z)=H(d)=1-H(c)$. 
Note that $H(0)=0$ and $H(u)=1$, where $u$ is a unit modulus complex number, 
because the initial states in these cases are described by one eigenvector. 
Born rule corresponds to $H(c)=|c|^2$.
We will apply the same generalized probability assignment rule to both Alice's and Bob's measurements, 
because there is no reason to distinguish Alice's and Bob's qubits and measurements. 

Here we consider the non-negative function $H(c)$ as a function $\mathcal{H}(|c|^2)$ of the absolute value 
squared $|c|^2$ for clear understanding of the essential context. 
The derivation for the general non-negative real function $H(c)$ of $c$ is given in Appendix.    
We will consider a pair of entangled qubits because 
this is a minimal case, which has nonlocal quantum correlations, to signal each other 
between two parties.  
We will show that Born rule is derived by requiring no-causal-order condition on the new quantum correlations 
generated by the generalized quantum probability assignment.  

Alice and Bob are supposed to share the following general state for a pair of entangled qubits described by  
\begin{eqnarray}
|\psi\rangle_{AB} &=& \alpha_1 |0\rangle_{x=0} |0\rangle_{y=0} + \alpha_2 |0\rangle_{x=0} |1\rangle_{y=0}  \\ \nonumber
&+& \alpha_3 |1\rangle_{x=0} |0\rangle_{y=0} + \alpha_4 |1\rangle_{x=0} |1\rangle_{y=0},
\end{eqnarray} 
where $|\alpha_1|^2+|\alpha_2|^2+|\alpha_3|^2+|\alpha_4|^2=1$ and 
$|0\rangle_{x=0}$, $|1\rangle_{x=0}$, $|0\rangle_{y=0}$, and $|1\rangle_{y=0}$ are 
the eigenvectors of Alice's input $x$ and Bob's input $y$, respectively. 
We will denote $|a\rangle_{x=0} |b\rangle_{y=0}$ simply as $|a\rangle_{0} |b\rangle_{0}$.
Let us first calculate the joint probability of outcomes $(0,0)$ for inputs $(0,0)$ in Alice-first measurement $P_A(00|00)$. 
The initial state $|\psi\rangle_{AB}$ is a 4-dimensional vector not a two-dimensional vector so that 
there seems to have a problem to apply the generalized quantum probability assignment, defined for 
a qubit, to Alice's measurement. 
However, we can consider the state $|\psi\rangle_{AB}$ as an effective tow-dimensional vector by factoring 
with the eigenvectors of Alice's input observable $x=0$ as 
\begin{eqnarray}
|\psi\rangle_{AB} = \sqrt{C} |0\rangle_{0} |Y_1\rangle_{0}  + 
\sqrt{D} |1\rangle_{0} |Y_2\rangle_{0}, 
\end{eqnarray}
where $ C=  |\alpha_1|^2 +|\alpha_2|^2$ and $D= |\alpha_3|^2 +|\alpha_4|^2$. 
$|Y_1\rangle_{0}=({\alpha_1 |0\rangle_{0}+ \alpha_2|1\rangle_{0}})/{\sqrt{C}}$ and 
$|Y_2\rangle_{0}= ({\alpha_3 |0\rangle_{0}+ \alpha_4|1\rangle_{0}})/{\sqrt{D}}$ are normalized states. 
Then the two vectors $|0\rangle_{0} |Y_1\rangle_{0}$ and $|1\rangle_{0} |Y_2\rangle_{0}$ are 
orthogonal and form a two-dimensional Hilbert space so that the state $|\psi\rangle_{AB}$ is a vector 
in this two-dimensional Hilbert space. 
Here to represent the state $|\psi\rangle_{AB}$, we used the eigenstates of Bob's input $y=0$, however, 
the above argument is independent on the choice of Bob's input. 
Hence Alice's measurement as the first measurement can be considered as a measurement on a 
vector in two-dimensional Hilbert space. 
After Alice's measurement, the state $|\psi\rangle_{AB}$ collapses either to 
$|0\rangle_{0} |Y_1\rangle_{0}$ or to $|1\rangle_{0} |Y_2\rangle_{0}$ by the quantum measurement postulate iii)
with the probabilities determined by the generalized probability assignment. 
That is, the state of Bob, after the measurement of Alice with input $x=0$ and outcome $0$, is projected to 
$|Y_1\rangle_{0}$ with probability $\mathcal{H}(|\sqrt{C}|^2)$. 
Then the probability of outcome $0$ for Bob's measurement on the state 
$|Y_1\rangle_{0}$ is determined by $\mathcal{H}(|\alpha_1/\sqrt{C}|^2)$, hence the joint probability of a pair of 
outcomes $(0,0)$ for a pair of inputs $(0,0)$ of Alice and Bob in the Alice-first measurement is obtained 
by the product of $\mathcal{H}(|\sqrt{C}|^2)$ and $\mathcal{H}(|\alpha_1/\sqrt{C}|^2)$, i.e., 
\begin{eqnarray}
\label{eq:PrAF}
P_A(00|00)=\mathcal{H}({|\alpha_1|^2 +|\alpha_2|^2})
\mathcal{H}\left(\frac{|\alpha_1|^2}{{|\alpha_1|^2 +|\alpha_2|^2}}\right).
\end{eqnarray}
Now let us consider Bob-first measurement, in which the following factorization of the state $|\psi\rangle_{AB} $ is useful,
\begin{eqnarray}
|\psi\rangle_{AB} = \sqrt{E} |X_1\rangle_{0} |0\rangle_{0}  + 
\sqrt{F} |X_2\rangle_{0} |1\rangle_{0}
\end{eqnarray}
with $ E=  |\alpha_1|^2 +|\alpha_3|^2$, $F= |\alpha_2|^2 +|\alpha_4|^2$,  
$|X_1\rangle_{0}=({\alpha_1 |0\rangle_{0}+ \alpha_3|1\rangle_{0}})/{\sqrt{E}}$,  
and $|X_2\rangle_{0}= ({\alpha_2 |0\rangle_{0}+ \alpha_4|1\rangle_{0}})/{\sqrt{F}}$.
By a similar calculation to Alice-first measurement, the joint probability of Bob-first measurement for 
the same inputs $(0,0)$ and outputs $(0,0)$ as the Alice-first measurement can be obtained as
\begin{eqnarray}
\label{eq:PrBF}
P_B(00|00) =\mathcal{H}({|\alpha_1|^2 +|\alpha_3|^2}) 
\mathcal{H}\left(\frac{|\alpha_1|^2}{{|\alpha_1|^2 +|\alpha_3|^2}}\right),
\end{eqnarray}
applying the same generalized probability assignment to Bob-first measurement.
 
We impose no-causal-order condition to the above results, which requires $P_A(00|00) = P_B(00|00)$. 
This condition should be satisfied for arbitrary $\alpha_1$, $\alpha_2$, $\alpha_3$, and $\alpha_4$. 
By substituting $0$ for $\alpha_3$, the equality of Eqs. (\ref{eq:PrAF}) and (\ref{eq:PrBF}) gives 
the following relation 
\begin{eqnarray}
\label{eq:NCO}
\mathcal{H}(|\alpha_1|^2 +|\alpha_2|^2)\mathcal{H}\left(\frac{|\alpha_1|^2}{|\alpha_1|^2 +|\alpha_2|^2}\right)
= \mathcal{H}(|\alpha_1|^2 ),
\end{eqnarray}
because $\mathcal{H}(1^2)=H(1)=1$.  
The above relation also has to be satisfied when $\alpha_1$ and $\alpha_2$ are exchanged with each other 
because of the freedom of relabeling outcomes $0\leftrightarrow 1$. 
And then we obtain the relation
\begin{eqnarray}
\label{eq:NDO}
\mathcal{H}(|\alpha_1|^2 +|\alpha_2|^2)\mathcal{H}\left(\frac{|\alpha_2|^2}{|\alpha_1|^2 +|\alpha_2|^2}\right)
= \mathcal{H}(|\alpha_2|^2 ).
\end{eqnarray} 
 By adding those two relations in Eqs. (\ref{eq:NCO}) and (\ref{eq:NDO}) we obtain 
\begin{eqnarray} \nonumber
\mathcal{H}(C)\left[\mathcal{H}\left(\frac{|\alpha_1|^2}{C}\right)
+ \mathcal{H}\left(\frac{|\alpha_2|^2}{C}\right)\right]   
= \mathcal{H}(|\alpha_1|^2 )+\mathcal{H}(|\alpha_2|^2 ).
\end{eqnarray}
The addition of two probabilities, $\mathcal{H}\left({|\alpha_1|^2}/({|\alpha_1|^2 +|\alpha_2|^2})\right)$ and 
$ \mathcal{H}\left({|\alpha_2|^2}/({|\alpha_1|^2 +|\alpha_2|^2})\right)$, becomes $1$ from the probability normalization 
because the sum of two arguments ${|\alpha_1|^2}/({|\alpha_1|^2 +|\alpha_2|^2})+
{|\alpha_2|^2}/({|\alpha_1|^2 +|\alpha_2|^2})$ is $1$. Finally we get the following relation 
\begin{eqnarray}
\mathcal{H}(|\alpha_1|^2 +|\alpha_2|^2)=\mathcal{H}(|\alpha_1|^2 )+\mathcal{H}(|\alpha_2|^2 ),
\end{eqnarray}
which requires that the functional form of $\mathcal{H}(|c|^2)$ should be linear. 
Considering the probability normalization, $\mathcal{H}(|c|^2)$ is determined as $\mathcal{H}(|c|^2)=|c|^2$, which is 
the Born rule.  
In consequence, we have proven that no-causal-order condition makes Born rule the only quantum probability assignment, 
consistent with no-causal-order condition, for qubit states. 
The proof in higher dimensional Hilbert space will be essentially the same 
because the Hilbert space of the minimal case can always be considered as the subspace of higher dimensional Hilbert space. 
Our derivation of Born rule implies that no post-quantum theory, preserving relativistic causality, 
can be obtained by generalizing probability assignment from Born rule. 
Since we have derived Born rule from the generalization of the quantum probability assignment to a function of an arbitrary 
complex coefficient $c$ in Appendix, 
this proof is general in the sense that we assumed most general probability assignment by 
an arbitrary non-negative function of an arbitrary complex number to quantum measurement.

As a reference, we briefly show that no-signaling condition cannot determine a specific form 
for the general quantum probability assignment $\mathcal{H}(|c|^2)$. 
To study no-signaling condition, we have to consider Alice's another input $x=1$. 
By using eigenvectors $|+\rangle_1$ and $|-\rangle_1$ of the observable $x=1$, the state $|\psi\rangle_{AB}$ 
is rewritten as 
\begin{eqnarray}
|\psi\rangle_{AB} &=& \beta_1 |+\rangle_{1} |0\rangle_{0} + \beta_2 |+\rangle_{1} |1\rangle_{0} \\ \nonumber
&+& \beta_3 |-\rangle_{1} |0\rangle_{0} + \beta_4 |-\rangle_{1} |1\rangle_{0},
\end{eqnarray} 
where $|\beta_1|^2+|\beta_2|^2+|\beta_3|^3+|\beta_4|^2=1$. 
Then the no-signaling condition $P_A(0|00)=P_A(0|10)$ gives the relation
\begin{eqnarray}
 \mathcal{H}(|\sqrt{C}|^2)H\left(\frac{|\alpha_1|^2}{|\sqrt{C}|^2}\right) + 
\mathcal{H}(|\sqrt{D}|^2)\mathcal{H}\left(\frac{|\alpha_3|^2}{|\sqrt{D}|^2}\right) &&\\ \nonumber
= \mathcal{H}(|\sqrt{K}|^2)H\left(\frac{|\beta_1|^2}{|\sqrt{K}|^2}\right) + 
\mathcal{H}(|\sqrt{L}|^2)\mathcal{H}\left(\frac{|\beta_3|^2}{|\sqrt{L}|^2}\right),&&
\end{eqnarray}
where $K=|\beta_1|^2+|\beta_2|^2$ and $L=|\beta_3|^2+|\beta_4|^2$. This relation cannot determine 
the form of $\mathcal{H}(|c|^2)$, however, one can check this relation holds for Born rule, i.e., 
$\mathcal{H}(|c|^2)=|c|^2$, 
because $\alpha_1|0\rangle_0 +\alpha_3|1\rangle_0= \beta_1|+\rangle_1+\beta_3|-\rangle_1$.

\section{Discussion}

In this paper, we have investigated whether a post-quantum theory with superquantum correlations, 
preserving relativistic causality, can be obtained by modifying the quantum theory to generalize 
quantum probability assignment from Born rule. 
We have defined the new condition of causality, no-causal-order condition, for space-like separated events 
and proved that no-causal-order condition is stronger than no-signal condition between a pair 
of nonlocal binary devices.

We have proven that Born rule is the only possible probability assignment to quantum measurement 
under relativistic causality. 
The pair of qubits shared by two space-like separate parties is minimal models to show 
nonlocal correlations, hence this model is enough to investigate the limit on the probability assignment 
of quantum measurement by relativity. Note that one can always choose two orthogonal vectors to use as a qubit, 
at least mathematically, in higher dimensional Hilbert space. 
We have shown that no-causal-order condition derives Born rule, but no-signaling condition cannot.  
This derivation provides the understanding how relativity limits the nonlocal correlations of the quantum theory 
described in Hilbert space through measurement probability assignment. 
By this work, we hope to give a hint to understand the question of ``Why is not quantum theory more nonlocal?''.

\section{appendix}
\subsection{Contradiction of bi-directional signaling}

\begin{figure}[ht]
\centering
\begin{minipage}[b]{0.45\linewidth}
\includegraphics[width=\linewidth]{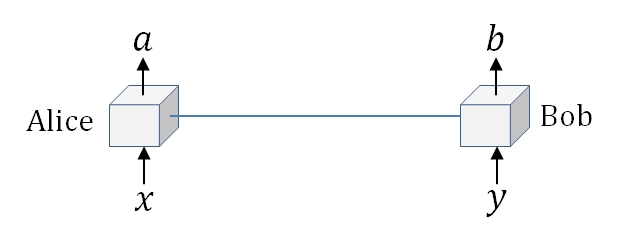}
\caption{A pair of nonlocal devices}
\label{fig:NLDevices}
\end{minipage}
\quad
\begin{minipage}[b]{0.45\linewidth}
\includegraphics[width=\linewidth]{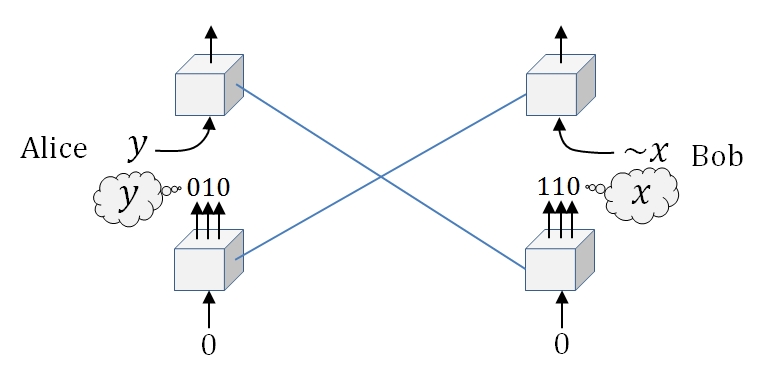}
\caption{The specific operation of two pairs of nonlocal devices which shows contradiction.}
\label{fig:CDOperation}
\end{minipage}
\end{figure}

We will show that bi-directional signaling in time provides a contradiction by using an example operation 
sketched in Fig. \ref{fig:CDOperation}. Alice and Bob share two pairs of nonlocally correlated binary devices, 
which are connected by line in Fig. \ref{fig:CDOperation}. 
We assume the vertical upward direction is the time direction toward future. 
We call first- and second-box, respectively, according to the time sequence of operations. 
Then Alice's first (second) Box is correlated with Bob's second (first) box. 
Let us assume Alice and Bob put the same inputs, e.g. $0$ here, to their first-boxes. 
By analyzing their own outputs of the first boxes they could know the inputs of their partner's second boxes 
because the nonlocal correlations of each pair of box are bi-direction in time. 
That is, the knowledge of inputs of future boxes can be signaled 
backward to the past through nonlocally correlated boxes. 
Let us assume that Alice and Bob are supposed to operate their second boxes according to the rule that 
the input of Alice's second box is the the same as the input of Bob's second-box known by the backward signaling 
and the Bob's input of his second box is the opposite to the input of Alice's second box also known by 
the backward signaling. 
 Fig. \ref{fig:CDOperation} shows this situation: 
Alice knows Bob's input $y$ of second box by analyzing her first box's outputs so 
that Alice put $y$ into her second box. On the other hand, Bob's analysis on his first box's outputs 
informs the input $x$ of Alice's first box so that Box put $\sim x=x+1$ modulo $2$ into his second box. 
In order that this situation has no contradiction, $x=y$ and $\sim x=y$ must be satisfied, which is definitely impossible. 

Hence we have shown that bi-directional signaling invoke a logical contradiction.

\subsection{ Derivation of Born rule for general probability assignment function $H(c)$}

We will prove that $H(c)=|c|^2$ by considering the no-causal-order condition of $P_A(00|00)=P_B(00|00)$. 
To consider $H(c)$ as a function of $c$ not of $|c|^2$, the initial state $|\psi\rangle_{AB}$ 
suitable for Alice-first measurement should be rewritten as
\begin{eqnarray}
|\psi\rangle_{AB}= c |0\rangle_0 |y_1\rangle_0 + d |1\rangle_0 |y_2\rangle_0,
\end{eqnarray}
where 
$c= \sqrt{|\alpha_1|^2 +|\alpha_2|^2}e^{i\phi}$, and $d = \sqrt{|\alpha_3|^2 + |\alpha_4|^2}e^{i\theta}$. 
The states $|y_1\rangle_0= ({\alpha_1 |0\rangle_{0}+ \alpha_2|1\rangle_{0}})/c$ and 
$|y_2\rangle_{0}= ({\alpha_3 |0\rangle_{0}+ \alpha_4|1\rangle_{0}})/d$ are easily checked 
to have unit norms. 
Then
\begin{eqnarray}
P_A(00|00)=H(c)H\left( \frac{\alpha_1}{c}\right).
\end{eqnarray}
The useful description of the initial state for Bob-first measurement is
\begin{eqnarray}
|\psi\rangle_{AB}= e |x_1\rangle_0 |0\rangle_0 + f |x_2\rangle_0 |1\rangle_0,
\end{eqnarray}
where 
$e= \sqrt{|\alpha_1|^2 +|\alpha_3|^2}e^{i \eta}$, and $f =\sqrt{ |\alpha_2|^2 + |\alpha_4|^2}e^{i\xi}$. 
The states $|x_1\rangle_0= ({\alpha_1 |0\rangle_{0}+ \alpha_3|1\rangle_{0}})/e$ and 
$|x_2\rangle_{0}= ({\alpha_2 |0\rangle_{0}+ \alpha_4|1\rangle_{0}})/f$ are normalized states. 
Then
\begin{eqnarray}
P_B(00|00)=H(e)H\left( \frac{\alpha_1}{e}\right).
\end{eqnarray}

If we let $\alpha_3=0$, the no-causal-order condition $P_A(00|00)=P_B(00|00)$ becomes  
\begin{eqnarray}
\label{eq:NCOCM}
H(c)H\left( \frac{\alpha_1}{c}\right)=H(|\alpha_1|e^{i\eta})H\left( \frac{\alpha_1}{|\alpha_1|e^{i\eta}}\right)
= H(|\alpha_1|e^{i\eta}),
\end{eqnarray}
where we have used $H\left(\alpha_1/(|\alpha_1| e^{i\eta})\right)=1$ because $H(u)=1$ for a uni-modular complex number $u$. 
Using $\alpha_1$ and $\alpha_2$ exchange symmetry, the relation in Eq. (\ref{eq:NCOCM}) becomes
\begin{eqnarray}
\label{eq:A1A2EX}
H(c)H\left( \frac{\alpha_2}{c}\right)= H(|\alpha_2|e^{i\tilde{\eta}}),
\end{eqnarray}
where the argument $\tilde{\eta}$ is defined similar to $\eta$. 
 
By addition of two Eqs. (\ref{eq:NCOCM}) and (\ref{eq:A1A2EX}), we obtain
\begin{eqnarray}
H(c) \left[H\left( \frac{\alpha_1}{c}\right)+  H\left( \frac{\alpha_2}{c}\right)\right]
= H(|\alpha_1|e^{i\eta}) +  H(|\alpha_2|e^{i\tilde{\eta}}).
\end{eqnarray} 
Because $H(\alpha_1/c) +H(\alpha_2/c)=1$, the above relation becomes 
\begin{eqnarray}
\label{eq:MEBR}
H(c)=H(\sqrt{|\alpha_1|^2 +|\alpha_2|^2}e^{i\phi})= H(|\alpha_1|e^{i\eta}) +  H(|\alpha_2|e^{i\tilde{\eta}}).
\end{eqnarray}
This equation must satisfy for arbitrary $\alpha_1$, $\alpha_2$, $\phi$, $\eta$, and $\tilde{\eta}$ so that 
by letting $\alpha_1=0$ we obtain 
\begin{eqnarray}
H(|\alpha_2|e^{i\phi})=H(|\alpha_2|e^{i\tilde{\eta}}).
\end{eqnarray}
To satisfy this relation for arbitrary $\phi$ and $\tilde{\eta}$, the function $H(c)$ of $c$ should not be a 
function of an argument of $c$, but the absolute value of $|c|$. 
Eq. (\ref{eq:MEBR}) requires that the functional form of $H(c)$ should be $|c|^2$, which is Born rule. Q.E.D

\section*{Acknowledgements }

This work was supported by the National Research Foundation of Korea Grant funded by the Korean Government 
(2014-0379).


\begin{thebibliography}{99}

{\sf

 
\bibitem{Bell} J. S. Bell, 
Physics \textbf{ 1}, 195 (1964).
\bibitem{Bancal} J-D. Bancal, S. Pironio, A. Acin, Y-C. Liang, V. Scarani, V. and N. Gisin,  
Nature Phys. \textbf{ 8}, 867 (2012). 
\bibitem{Salart} D. Salart, A. Baas, C. Branciard, N. Gisin N. and H. Zbinden,  
Nature \textbf{ 454}, 861 (2008).
 \bibitem{Yin} J. Yin, Y. Cao, H-L. Yong, J-G. Ren, H. Liang, S-K. Liao, F. Zhou, C. Liu, Y-P. Wu, G-S. Pan, 
Q. Zhang, C-Z. Peng, and J-W. Pan,  
	Phys. Rev. Lett. \textbf{110}, 260407 (2013). 
     \bibitem{Clauser} J. F. Clauser, M. A. Horne, A. Shimony A. and R. A. Holt,  
		Phys. Rev. Lett. \textbf{ 23}, 880 (1969).
 \bibitem{Tsirelson} B. S. Tsirelson,  
Lett. Math. Phys. \textbf{ 4}, 93 (1980).
 \bibitem{Popescue1} S. Popescu and D. Rohrlich,  
Found. Phys. \textbf{ 24}, 379 (1994).
\bibitem{Brassard} G. Brassard, H. Buhrman, N. Linden, A. A. Methot, A. Tapp, A. and F. Unger,  
Phys. Rev. Lett. \textbf{96}, 250401 (2006).
\bibitem{Linden} N. Linden, S. Popescu, A. J. Short and A. Winter, 
 Phys. Rev. Lett. \textbf{99}, 180502 (2007).
\bibitem{Pawlowski} M. Pawlowski, T. Paterek, D. Kaszlikowski, V. Scarani, A. Winter and M. Zukowski, 
Nature \textbf{461}, 1101 (2009).
\bibitem{Navascues} M. Navascu\'es and H. Wunderlich,
Proc. Royal Soc. A \textbf{466} 881-890 (2009).
\bibitem{Fritz} T. Fritz, A. B. Sainz, R. Augusiak, J. B. Brask, R. Chaves, A. Leverrier and 
A. Ac\'in,  
Nat. Commun. \textbf{4} 2263 (2013).
\bibitem{Navascues1} M. Navascu\'es, Y. Guryanova, M. J. Hoban and A. Ac\'in,  
Nat. Commun. \textbf{6} 6288 (2015).
\bibitem{Wilde} M. M. Wilde, {\textit Quantum Information Thoery} (Cambridge University Press, New York, 2013).
}

\end{thebibliography}





\end{document}